\documentclass[aps,prl,reprint,superscriptaddress,showpacs]{revtex4-1}

\usepackage{color}

\usepackage{graphicx}
\usepackage{booktabs}
\usepackage{bm}
\usepackage{longtable}
\usepackage{amsmath}
\usepackage{dcolumn}
\usepackage{placeins}
\usepackage{expdlist}

\usepackage{siunitx}

\begin{document}

\title{Enhanced material defects imaging  with a radio-frequency atomic magnetometer}

\author{P. Bevington} 
\affiliation{National Physical Laboratory, Hampton Road, Teddington, TW11 0LW, United Kingdom}
\affiliation{Department of Physics, University of Strathclyde, Glasgow G4 0NG, United Kingdom}
\author{R. Gartman} 
\affiliation{National Physical Laboratory, Hampton Road, Teddington, TW11 0LW, United Kingdom}
\author{W. Chalupczak}
\affiliation{National Physical Laboratory, Hampton Road, Teddington, TW11 0LW, United Kingdom}

\date{\today}

\begin{abstract}
Imaging of structural defects in a material can be realized with a radio-frequency atomic magnetometer by monitoring the material's response to a radio-frequency excitation field. We demonstrate two measurement configurations that enable the increase of the amplitude and phase contrast in images that represent a structural defect in highly electrically conductive and high magnetic permeability samples. Both concepts involve the elimination of the excitation field component, orthogonal to the sample surface, from the atomic magnetometer signal. The first method relies on the implementation of a set of coils that directly compensates the excitation field component in the magnetometer signal. The second takes advantage of the fact that the radio-frequency magnetometer is not sensitive to the magnetic field oscillating along one of its axes. Results from simple modelling confirm the experimental observation and are discussed in detail.
\end{abstract}

%OCIS codes: (280.4788) Optical sensing and sensors; (110.0110) Imaging systems; (110.6955) Tomographic imaging.

\maketitle

%\textbf{Introduction.} 
Implementation of radio-frequency (rf) magnetic fields in non-destructive testing provides a cost-effective option for the detection of structural defects, particularly in cases when there is no direct access to the surface of the studied sample. The technique relies on monitoring the material response to the so-called primary magnetic field ($\vec{B}$) created by an rf coil \cite{Ma2017}. The material response can be detected in a variety of ways. Traditionally this is achieved by monitoring the impedance of the rf coil (or a dedicated pickup coil) \cite{Ma2017, Griffiths2001, Auld1999, Perez2004, Sophian2017}. However, the simplicity of instrumentation in this type of measurement is outweighed by the signal sensitivity degradation at low frequencies. Alternatives involve the implementation of magnetic sensors such as giant magnetoresistance (GMR) magnetometers \cite{Dogaru2000, Dogaru2001, Ripka2010},  superconducting quantum interference devices (SQUIDs) \cite{Krause2002, Storm2017}, and radio-frequency magnetometers \cite{Wickenbrock2014, Deans2016, Wickenbrock2016, Deans2017, Bevington2018a}. The magnetic field sensors  directly monitor the response of the medium, the so-called secondary magnetic field ($\vec{b}$). The secondary field is produced by the primary magnetic field through eddy currents excited in highly electrically conductive samples, or magnetisation induced in samples with a high magnetic permeability \cite{Bevington2018a}, and contains signatures of the inhomogeneities/ structural defects within the sample.

The focus of our studies is material defect imaging with a rf atomic magnetometer. We have previously demonstrated detection of defects (in the form of a recess) in carbon steel objects \cite{Bevington2018a} and analysed the shape of the spatial profiles observed in amplitude and phase images \cite{Bevington2018b}. The images recorded are produced from the variations in the phase and amplitude of the rf spectroscopy signal measured by the atomic magnetometer while scanning across the material. The rf field monitored by the magnetometer contains contributions from the primary and secondary magnetic field. A strong primary field contribution in the magnetometer signal results in the mapping of the two orthogonal components of the secondary field, $b_z$ and $b_y$, onto the amplitude and phase of the rf spectroscopy signal respectively.
We have previously reported that the component of the secondary field, $b_y$, parallel to the sample surface changes its sign in the vicinity of the surface crack (recess) \cite{Bevington2018b}. As a consequence of a strong primary field, variations in the resultant field recorded by the rf atomic magnetometer measure the direction flip of the secondary field component, however the observed rf signal phase change is smaller than the actual change in $b_y$.

\begin{figure}[htbp]
\includegraphics[width=\columnwidth]{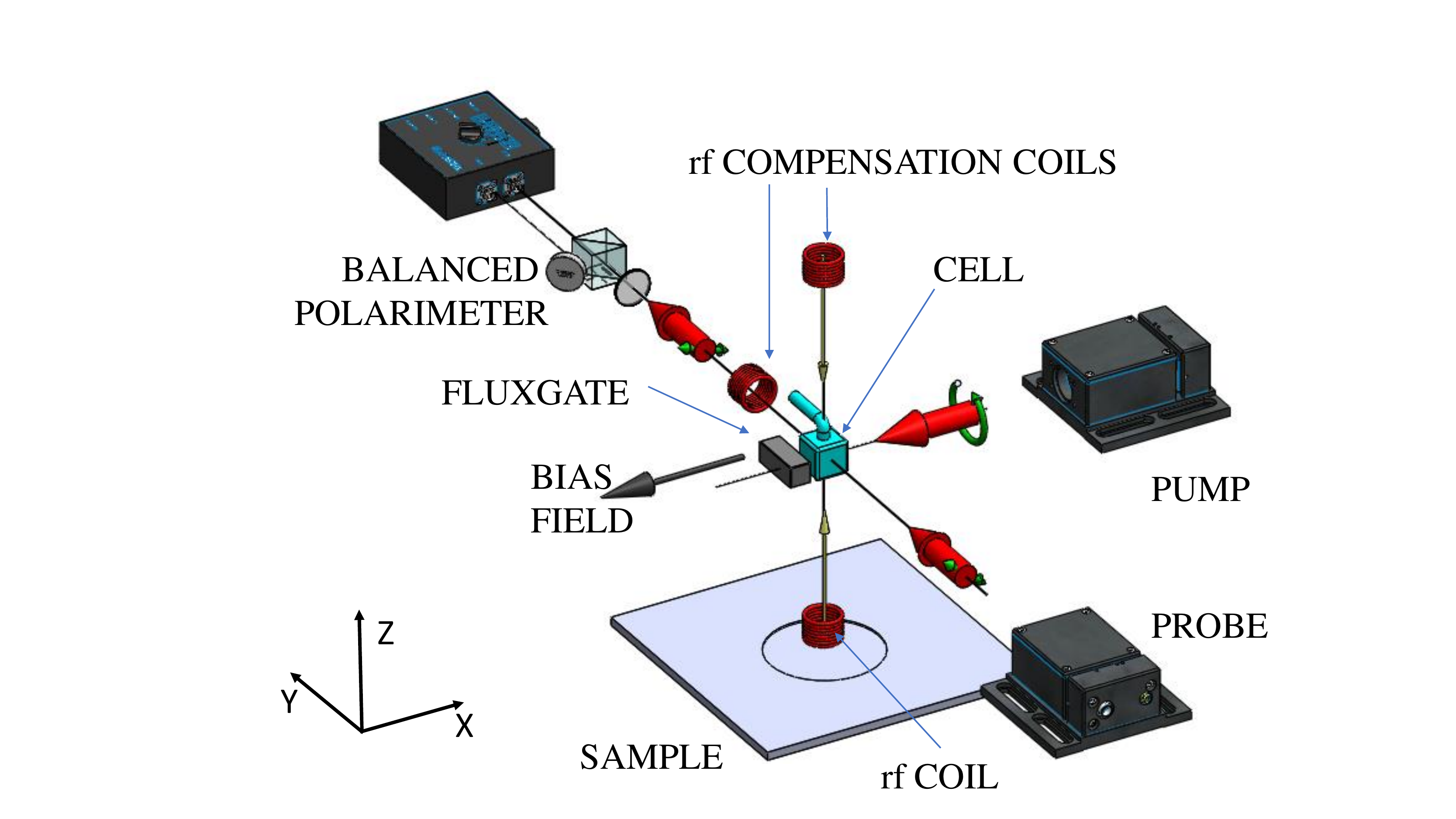}
\caption{Main components of the experimental setup. The secondary magnetic field is produced by the sample (plate with recess) response to the primary field created by the rf coil.  The atomic magnetometer signal combines components created by the primary field and the secondary magnetic field. The primary field contribution is compensated in the vapour cell by a set of two rf coils oriented along the $\hat{z}$ and $\hat{y}$ directions.}\label{fig:Setup}
\end{figure} 
 
In this paper, we demonstrate two simple methods of improving  structural defect imaging with an rf atomic magnetometer. Both concepts involve the elimination of the primary field component, orthogonal to the sample surface, from the atomic magnetometer signal. The first relies on the implementation of a set of coils  that creates an oscillating magnetic field, $\vec{B_C}$, that compensates the primary field contribution to the resultant field monitored by the atomic magnetometer vapour cell (Fig. \ref{fig:Setup}), without changing the efficiency of the $\vec{b}$ excitation. The second method benefits from the fact that the rf atomic magnetometer is not sensitive to magnetic fields oscillating along the static bias field axis, $\vec{B}_{bias}$ (Fig. \ref{fig:Setup}) \cite{Footnote1}. For $\vec{B}_{bias}$ oriented along $\vec{B}$ direction, $B_z$ will be absent in the rf atomic magnetometer signal and the measurement configuration becomes equivalent to that when the primary field component is compensated. We are going to show that the elimination of the primary field component from the magnetometer signal significantly increases the phase, and amplitude contrast of the images. This could provide a significatly quicker indication of a defect in non-destructive tests of large area samples. The concepts will be explored in the context of highly electrically conductive (aluminium) and highly magnetically permeable samples (carbon steel).

%\textbf{Experimental setup.} 
Since the experimental setup is similar to that described in \cite{Chalupczak2012, Bevington2018a, Bevington2018b} only the essential components will be briefly discussed here. The measurement signal comes from the phase and amplitude change in the rf resonance spectra registered by an rf atomic magnetometer as a sample is moved under the rf coil (Fig. \ref{fig:Setup}).
The rf coil producing $\vec{B}$ is driven by the internally referenced output of a lock-in amplifier (SRS 865). The samples are fixed to a 2D, computer-controlled translation stage. The sample is located approximately $\SI{30}{\centi\meter}$ from the cell and the coil is placed $\SI{1}{\milli\meter}$-$\SI{2}{\milli\meter}$ above the sample surface. A set of two rf coils oriented along $\hat{z}$ and $\hat{y}$ creates an oscillating magnetic field, $\vec{B_C}$, that compensates the primary field seen by the atomic magnetometer. The compensation coils are driven by the output of the same lock-in amplifier used to generate $\vec{B}$. This keeps a constant phase difference between the fields  $\vec{B}$ and  $\vec{B_C}$. The amplitudes of the two components of $\vec{B_C}$ are varied by changing the distance of the respective coils from the vapour cell. The resultant magnetic field monitored by the rf atomic magnetometer includes components from the primary, secondary and compensation field, i.e. $\vec{b}+\vec{B}+\vec{B_C}=\vec{b}+\vec{B'}$. 
The measurements are performed in a magnetically unshielded environment, where $\vec{B}_\mathrm{bias}$ is stabilised by three pairs of nested, orthogonal, square Helmholtz coils \cite{Deans2017, Bevington2018a, Bevilacqua2016}. The strength of the bias field ($\lvert \vec{B}_\mathrm{bias} \rvert$) defines the operating frequency of the system (in most of the measurements equal to $\SI{12.6}{\kilo\hertz}$), i.e. the frequency of the magnetic resonance and the required primary field frequency.
The atomic magnetometer is based on room temperature caesium atomic vapour (atomic density $n_{\text{Cs}}=\SI{3.3e10}{\per\centi\meter\cubed})$ housed in a paraffin-coated glass cell. Caesium atoms are optically pumped into the stretched state ($F= 4, m_{F} = 4$) with a circularly polarised laser locked to the Cs $6\,^2$S$_{1/2}$ F=3$\rightarrow{}6\,^2$P$_{3/2}$ F'=2 transition (D2 line, $\SI{852}{\nano\meter}$) propagating along the bias magnetic field $\vec{B}_{bias}$. The probe beam ($\SI{30}{\micro\watt}$) is $\SI{580}{\mega\hertz}$ blue shifted from the $6\,^2$S$_{1/2}$ F=4$\rightarrow{}6\,^2$P$_{3/2}$ F'=5 transition via phase-offset-locking to the pump beam. Coherent spin precession of the Cs atoms is coupled to the polarisation of the probe beam (Faraday rotation) and is detected with a balanced polarimeter. 
 Our experimental configuration, Fig. \ref{fig:Setup}, allows the determination of amplitude and phase (orientation) of the rf field in the $yz$ plane, since only the components perpendicular to bias magnetic field excite atomic precession. The amplitude (R=$\sqrt{(b_y+B'_y)^2+( b_z+B'_z)^2}$) and the phase ($\phi = \arctan\frac{ b_y+B'_y}{ b_z+B'_z}$) of the rf spectroscopy signal describe the changes in the resultant field, where $ b_z+B'_z$ and $ b_y+B'_y$ are the two quadrature components of the rf signal.

In most of the measurements, the frequency of the primary field has been scanned across the rf resonance, i.e. the whole resonance profile has been recorded, for each point of the image \cite{Bevington2018a}. We also tested an alternative mode of data acquisition, which significantly decreases the image acquisition time. In this mode, the modulation of the $\vec{B}$ frequency has been replaced with low-frequency modulation ($\SI{1}{\hertz}$-$\SI{20}{\hertz}$) of the amplitude of $\vec{B}_{bias}$. 
Since the amplitude of the bias field is stabilised with the fluxgate magnetometer \cite{Bevington2018a}, a small planar coil placed on top of the fluxgate magnetometer is used to modulate $\vec{B}_{bias}$. As a result of this, the output of the fluxgate contains an oscillatory component in the relevant direction. In this case, the signal recorded by a lock-in amplifier demodulating the magnetometer signal at the primary field frequency will contain a low-frequency oscillation with an amplitude equal to that of the rf resonance amplitude. A second lock-in amplifier referenced to the frequency of the $\vec{B}_{bias}$ amplitude modulation enables the readout of the rf resonance amplitude. The extent of the $\vec{B}_{bias}$ amplitude modulation can accommodate any imperfection in the bias field stabilisation, i.e. possible shifts in resonance frequency at different locations across the sample.

\begin{figure}[bt]
\includegraphics[width=\columnwidth]{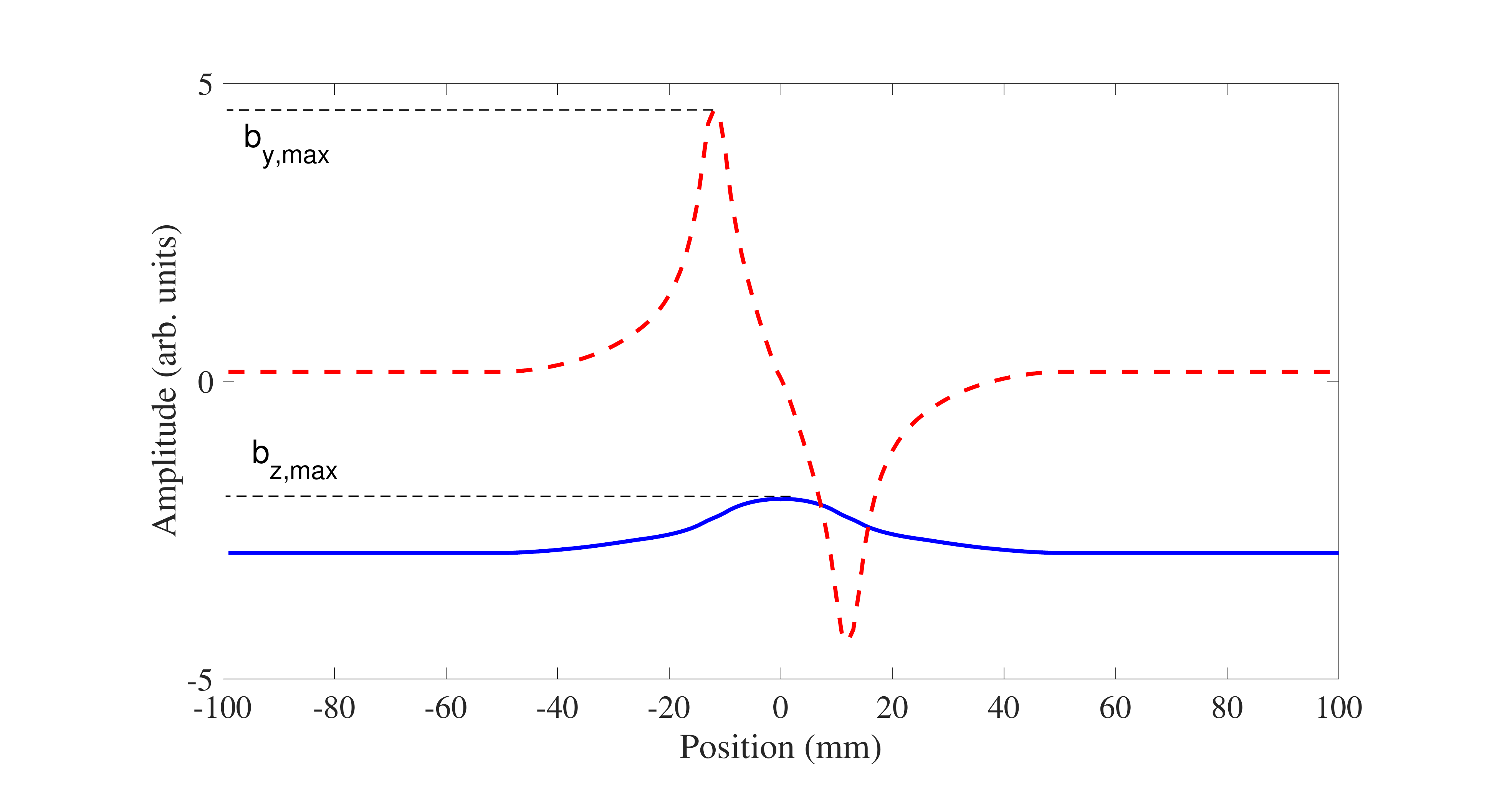}
\caption{Simulation of the secondary field components $b_y$ and $b_z$ (marked with dashed red and solid blue lines respectively). The coil is moved along the $y$ axis across the centre of the recess. Maximum values of $ b_y$, and $ b_z$ within the recess boundaries are referred to as $b_\mathrm{y, max}$, and $b_\mathrm{z, max}$.}\label{fig:BxByBz}
\end{figure}

\begin{figure*}[tb]
\includegraphics[width=\textwidth]{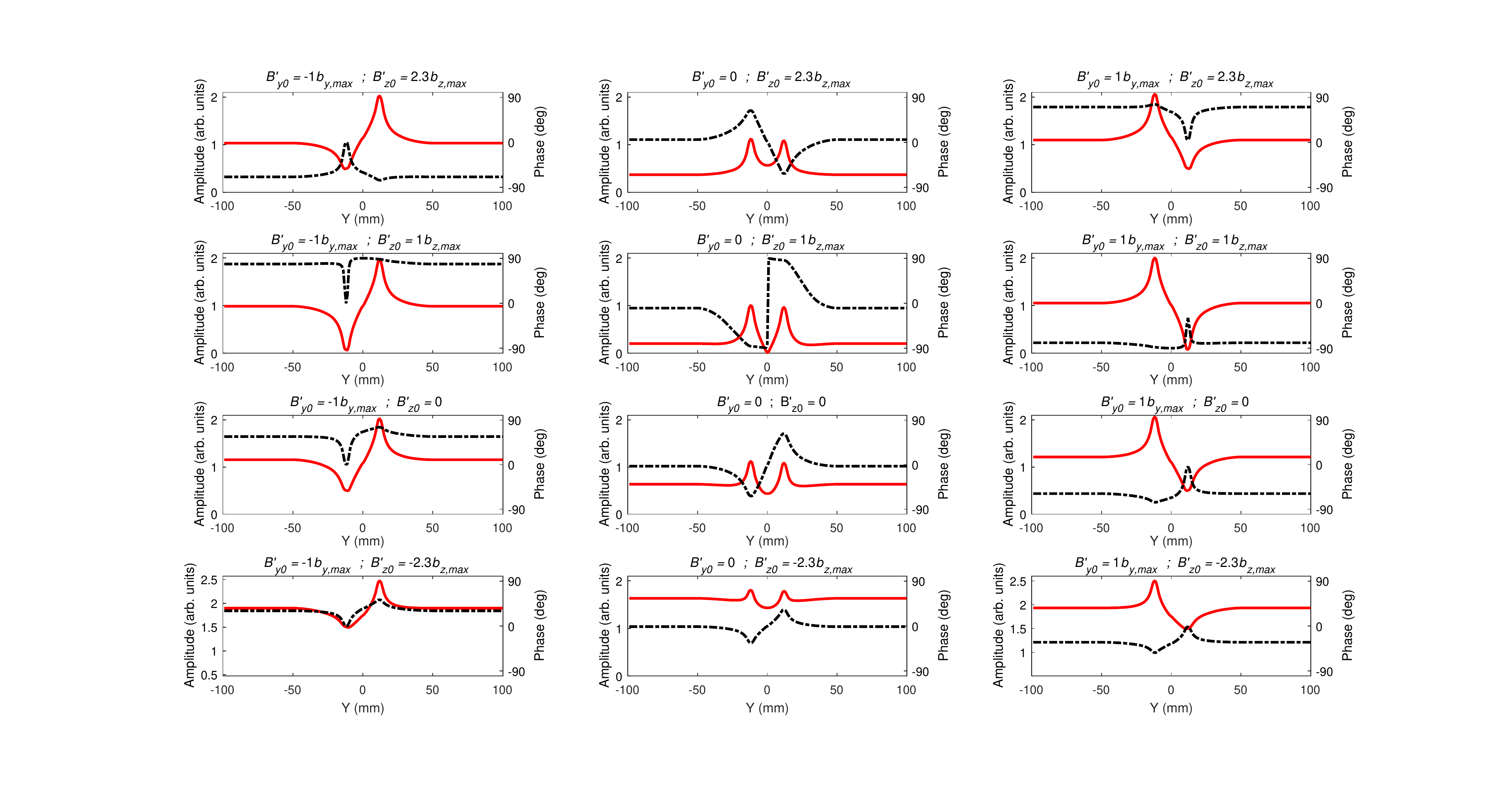}
\caption{The modelled change in the signal phase (dotted black line) and the amplitude (solid red line) of the magnetic resonance signal over a recess recorded by a magnetometer for various amplitudes of the primary field components. The vertical axis of the image array represents changes in the vertical component, while the horizontal axis represents changes in the horizontal component of the primary field. Amplitude expressed in units of $b_\mathrm{y, max}$.} \label{fig:Model1}
\end{figure*}

%\textbf{Modeling.}
We begin with a model of $\vec{b}$ in the case of a 1D scan of the rf coil position in the $\hat{y}$ direction across the centre of the recess for a highly electrically conductive sample, Fig. \ref{fig:Setup} \cite{Footnote2}. In this geometry only the $b_y$ and $b_z$ components have a non-zero value. Figure \ref{fig:BxByBz} shows the dependence of the secondary rf field components on the rf coil position. In a homogeneous sample (i.e. far from the recess), the eddy current flow has a circular symmetry and $\vec{b}$ has only one non-zero component, $b_z$ (solid blue line in Fig. \ref{fig:BxByBz}). In the vicinity of the recess, the symmetry of the eddy current flow is broken and a non-zero component of $\vec{b}$ is produced in the $yz$ plane (dashed red line in Fig. \ref{fig:BxByBz}). The asymmetry of the eddy current flow is mirrored on the other side of the recess, resulting in the opposite sign of $b_y$. In the case where the rf coil is above the recess, the magnetic flux through the plate becomes smaller since the coil lift-off is greater, hence $\lvert \vec{b} \rvert$ decreases. We will refer to the maximum value of $ b_y$, and $ b_z$ within the recess boundaries as $b_\mathrm{y, max}$, and $b_\mathrm{z, max}$.

As we have shown previously \cite{Bevington2018b}, in the presence of a strong primary field $\vec{B}$ along $\hat{z}$, changes in the secondary field components are mapped onto the amplitude ($b_z$) and phase ($b_y$) of the rf spectroscopy signal. Changes in the resultant magnetic field monitored by the atomic magnetometer are relatively small as they appear on top of a much bigger primary field.
To provide better insight, we consider the case where $\lvert \vec{b} \rvert \sim \frac{1}{3} \lvert \vec{B} \rvert$ along both $\hat z$ and $\hat y$. Here, the observed amplitude and phase contrasts are estimated to be $C_R \le 0.05$ and $C_\phi \le 4 ^{\circ}$ respectively, where $C_\mathrm{R} = (R_\mathrm{Max} - R_\mathrm{Min})/(R_\mathrm{Max} + R_\mathrm{Min})$ and $C_\mathrm{\phi} = \phi_\mathrm{Max} - \phi_\mathrm{Min}$, with $R_\mathrm{Max}, R_\mathrm{Min}, \phi_\mathrm{Max}$, and $\phi_\mathrm{Min}$ being max/ min values of relevant variables. These values are significantly lower than the theoretical maximum values of $C_\mathrm{R}=1$ and $C_\mathrm{\phi} = 180^{\circ}$.

Figure \ref{fig:Model1} shows the simulated dependence of the amplitude and phase of the rf spectroscopy signal measured by the atomic magnetometer on the resultant magnetic field. Moving along rows/ columns of Fig. \ref{fig:Model1} is equivalent to a variation in the $\hat y$/ $\hat z$ component of the resultant magnetic field. Firstly, we consider changes in the signal profile with $B'_z$ for $B'_y = 0$, i.e. plots in the middle column of Fig. \ref{fig:Model1}. Complete compensation of the primary field component in the resultant magnetic field monitored by the atomic magnetometer ($B'_y=0, B'_z=0$) leads to an increase in amplitude and phase contrast ($C_R = 0.44$ and $C_\phi = 124^{\circ}$, central plot in third row of Fig. \ref{fig:Model1}) in respect to previously quoted values, but still does not produce maximum contrast values. The reason is that for $\lvert \vec{B'} \rvert = 0$ the signal amplitude becomes R=$\sqrt{b_y^2+b_z^2}$, where the high contrast component $b_{y}$ is reduced by the slowly varying offset of $b_{z}$ (Fig. \ref{fig:BxByBz}). The phase of the rf spectroscopy signal ($\phi$) changes by $180^{\circ}$, only when there is a change of the sign in the numerator and a singularity in the denominator. This indicates that the condition for achieving maximum contrast, with respect to the $\hat z$ axis components, is when $B'_z + b_{z, max} = 0$. In the following, we refer to this condition, along with $B'_y = 0$,  as the compensation point. Fig. \ref{fig:Model1} confirms that the maximum amplitude and phase contrast is observed for $B'_z + b_\mathrm{z, max} = 0$, in each column, i.e. for an arbitrary value of $B'_y$. 

Moving right/ left from the compensation point amounts to the addition of a negative/ positive reference level. This induces a symmetry change of the profiles observed on either side of the compensation point. The amplitude and phase contrasts ($C_R = 1$,  $C_\phi = 180^{\circ}$) do not deteriorate in the vicinity of the compensation point as long as $|B'_y| \leq b_{y, max}$.
In the vicinity of the compensation point, modelling predicts phase jumps in the magnetic resonance signal by nearly $180^{\circ}$ over the recess area. The reason for the sudden phase change is the presence of a z component in the resultant rf field in the denominator of the arctan function that defines the phase of rf spectroscopy signal.

\begin{figure}[htbp]
\includegraphics[width=\columnwidth]{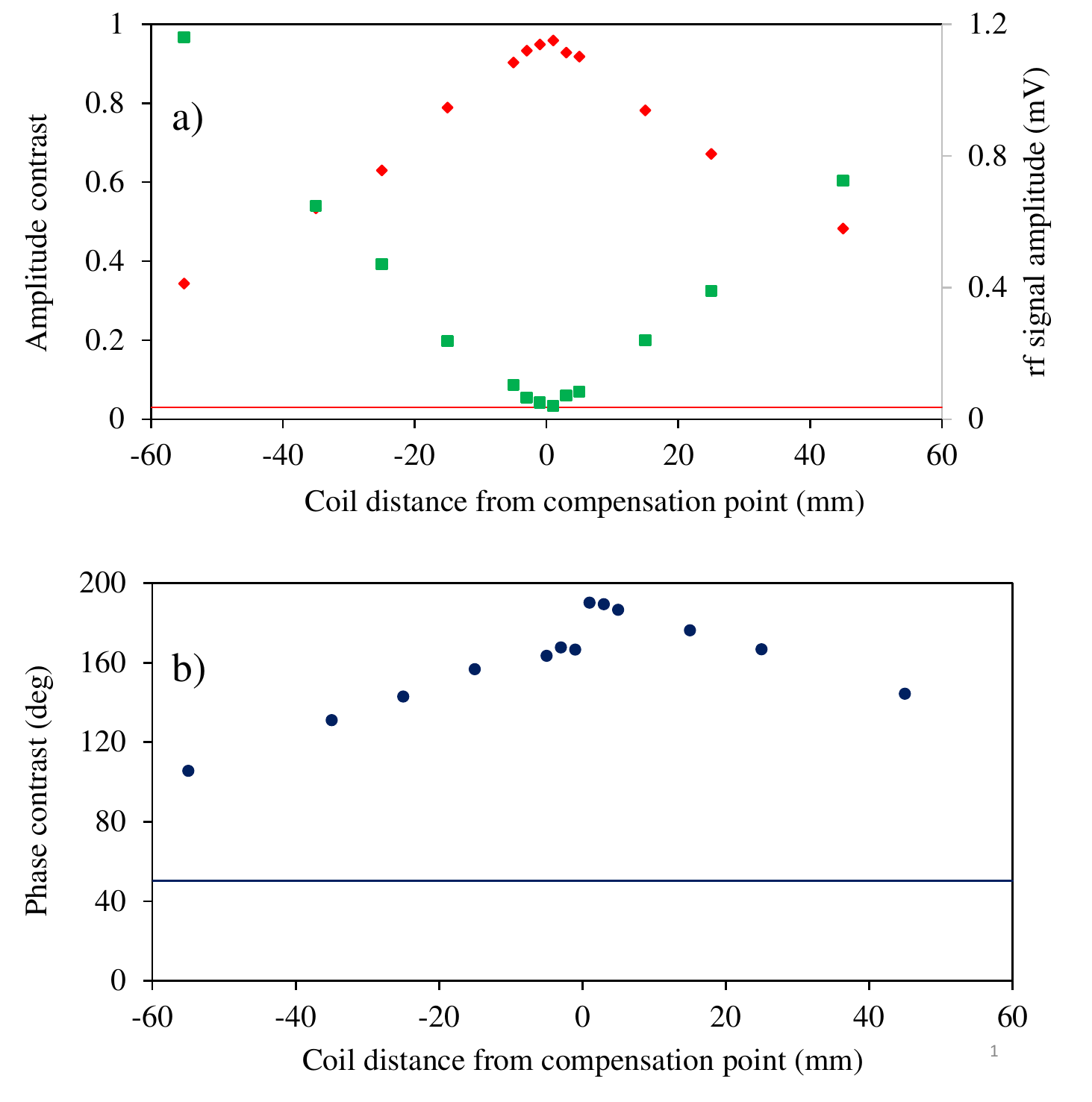}
\caption{(a) Amplitude and (b) phase contrast measured for different locations of the compensation coil above the vapour cell from the compensation point. Green squares in (a) represent the change of the rf spectroscopy signal amplitude. Solid red/ blue  line shows the amplitude/ phase contrast in the absence of a compensation field.} \label{fig:Contrast}
\end{figure}

%\textbf{Experimental results.} 
We adapt as a testbed for the experimental exploration of rf magnetic field compensation, changes in the amplitude and phase of the rf spectroscopy signal recorded with an rf coil scanned across defect in the form of recesses ($\SI{24.5}{\milli\meter}$ diameter, \SI{2.4}{\milli\meter} deep) in \SI{6}{\milli\meter} thick plate of aluminium and carbon steel \cite{Bevington2018a, Bevington2018b}. 
We begin our studies from the realization of the compensation point. Experimentally, this is achieved by tuning the distance between the compensation coils and the vapour cell (Fig. \ref{fig:Setup}). The coil located above the vapour cell is positioned on axis with the rf coil producing the primary field. The optimum location of this coil along $\hat{z}$ is established by minimising the amplitude of the rf spectroscopy signal over the recess. The position of the other compensation coil can be adjusted in all three directions \cite{Footnote3}. The presence of this coil is particularly important in measurements with objects with a high magnetic permeability, where a misalignment of the rf coil with respect to the surface normal of the sample results in significant $B_y, B_x$ components of the primary field. Figure \ref{fig:Model1} indicates that compensation in the horizontal direction results in symmetric amplitude and phase profiles. This factor is utilised in searches for the compensation point.

%\textbf{Contrast} 
Figure \ref{fig:Contrast} shows the changes in (a) amplitude, $C_R$ (red diamonds), and (b) phase, $C_\phi$, contrast as a function of the distance of the vertical coil from the compensation point  along $\hat {z}$. The measurement was performed with the aluminium plate. Green squares in Fig. \ref{fig:Contrast} (a) represent the change of the rf spectroscopy signal amplitude measured in the centre of the recess. Both plots confirm the the contrast is maximised at the compensation point. For reference, we show the amplitude/ phase contrast value recorded without rf compensation fields [red/ blue solid horizontal line in Fig. \ref{fig:Contrast} (a)/ (b)].

\begin{figure*}[tb]
\includegraphics[width=\textwidth]{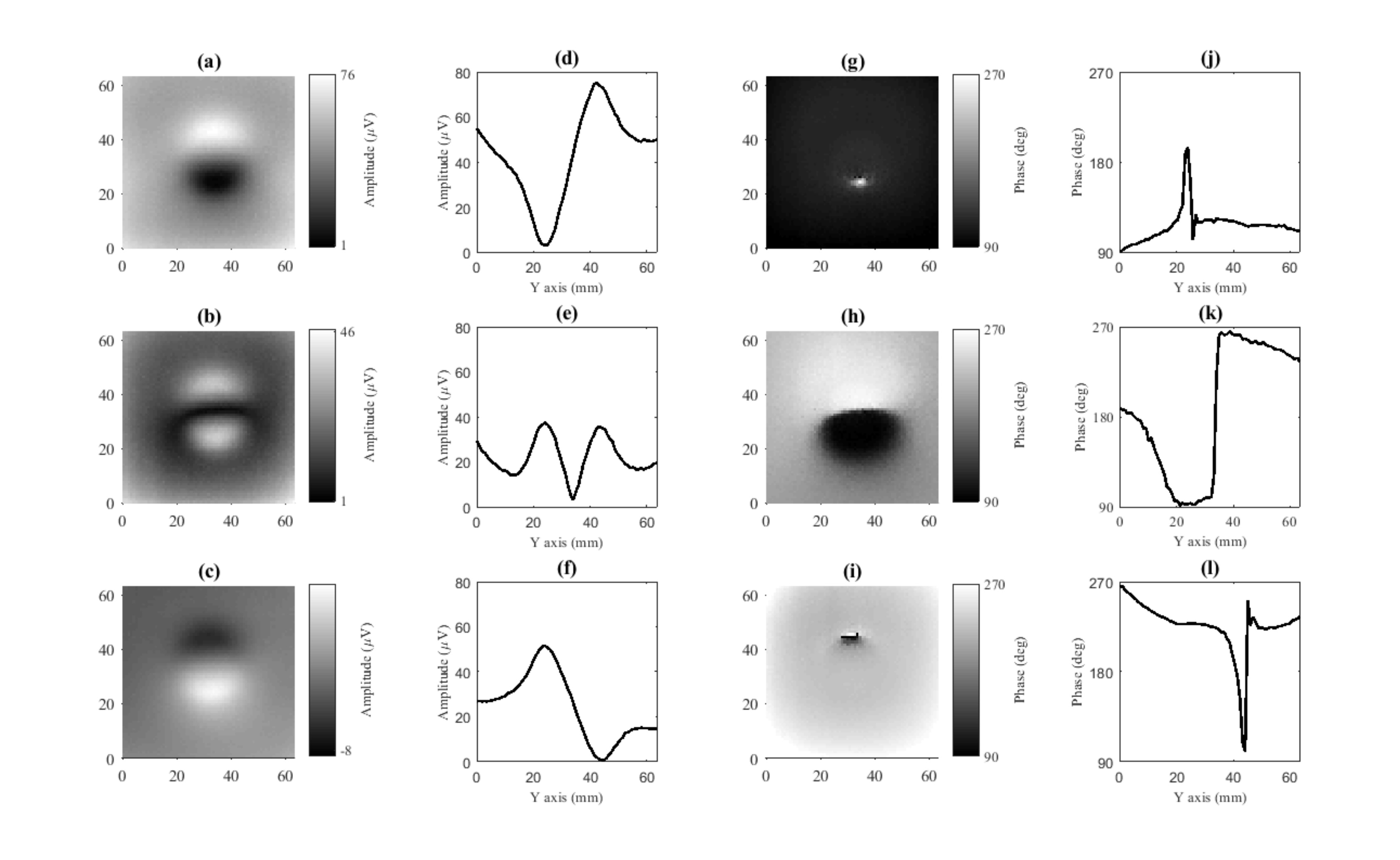}
\caption{(a)-(c)/ (g) - (i) The measured change of the amplitude/ phase of the rf spectroscopy signal over a $64\times\SI{64}{\milli\meter\squared}$ area of a \SI{6}{\milli\meter} thick carbon steel plate, containing a \SI{24}{\milli\meter} diameter recess that is \SI{2.4}{\milli\meter} deep, recorded for three values of the horizontal component of the compensation rf field around compensation point. (d)-(f)/ (j)-(l) The vertical cross-sections through the centre of the amplitude/ phases images. The images have been recorded at $\SI{114.2}{\kilo\hertz}$. } \label{fig:Map}
\end{figure*}

%\textbf{Shape} 
Figure \ref{fig:Map} (a)-(c)/ (g)-(i) shows the amplitude/ phase images of $64\times\SI{64}{\milli\meter\squared}$ area of a \SI{6}{\milli\meter} thick carbon steel plate containing a \SI{24.5}{\milli\meter} diameter recess that is \SI{2.4}{\milli\meter} deep recorded for three values of the horizontal component of the compensation rf field. The vertical cross-sections through the centre of the amplitude/ phase images are shown in Fig. \ref{fig:Map} (d)-(f)/ (j)-(l). They are equivalent to the calculated amplitude profiles from the second row of Fig. \ref{fig:Model1}. The cases illustrated in Fig. \ref{fig:Map} (b)/(e), ((h)/ (k) represent the compensation point. 
Near the compensation point, the amplitude image produced by a recess is symmetrical, i.e. two maxima corresponding to rising and falling edges. This results from zeroing the primary field that sets the background reference for changes in the secondary field generated by the magnetisation ($R \sim \lvert {b_y} \rvert $). 
The bright maximum in Fig. \ref{fig:Map} (b) is surrounded by a dark ring, in other words, the profiles created by the two recess edges are not perfectly symmetrical. This might be due to imperfect compensation of the static magnetic field, in addition to the rf fields. Imperfect compensation of the static magnetic field, i.e. a variation of the ambient magnetic field, is equivalent to a change in the direction of the atomic magnetometer axis, which could affect the observed image [see following section, Fig. \ref{fig:Images} (b) and (c)]. This could be minimised by increasing the operating frequency. The magnetic response of the sample can be thought of in terms of changes in the rf coil inductance. Inhomogeneity in the magnetisation across the plate would be seen as a variation of the primary field strength, meaning rf field compensation is valid only locally. 
Change of the rf field strength monitored by the magnetometer can be seen by the variation in the background level. There is a change in the shape and symmetry of the amplitude profiles on either side of the compensation point, related to the passage from R=$ \lvert b_y+B'_y \rvert$ [Fig. \ref{fig:Map} (d)] through $R \sim \lvert {b_y} \rvert $ [[Fig. \ref{fig:Map} (e)]] to $R=\lvert b_y-B'_y \rvert$ [Fig. \ref{fig:Map} (f)].
The phase contrast decreases as we move away from the compensation point, Fig. \ref{fig:Map} (g)-(i). 
Importantly, the area where the phase departs from its background value decreases away from the compensation point. This might have practical consequences in a measurement with a coarse spatial step, since the phase change generated by the recess could be non-visible as it would be equivalent to spatial integration of the phase change over a larger area.

%\textbf{Lift-off} 
The benefit of rf compensation can be demonstrated in the experiment with an increased lift-off distance, $\SI{6}{\milli\meter}$ - $\SI{7}{\milli\meter}$. The starting point for this is the measurement of the phase contrast in the standard configuration (the rf coil producing the primary field located $\SI{1}{\milli\meter}$ - $\SI{2}{\milli\meter}$ above the sample surface, no rf compensation). The phase contrast for a case of a \SI{12}{\milli\meter} diameter recess that is \SI{2.4}{\milli\meter} deep in a \SI{6}{\milli\meter} thick aluminium plate is $C_\phi =40^{\circ}$. An increase of the lift-off ($\SI{6}{\milli\meter}$ - $\SI{7}{\milli\meter}$) results in a reduction of the strength of the primary field at the sample, and consequently, the secondary field. At the same time, for a fixed distance between the vapour cell and the sample, Fig. \ref{fig:Setup}, the primary field component monitored by the atomic magnetometer increases. This causes a reduction of the recorded phase contrast to $C_\phi =20^{\circ}$. With the addition of the compensation rf field we were not only able to recover the initial phase contrast value but even increase it to its maximum value, $C_\phi =180^{\circ}$. 

\begin{figure}[htbp]
\includegraphics[width=\columnwidth]{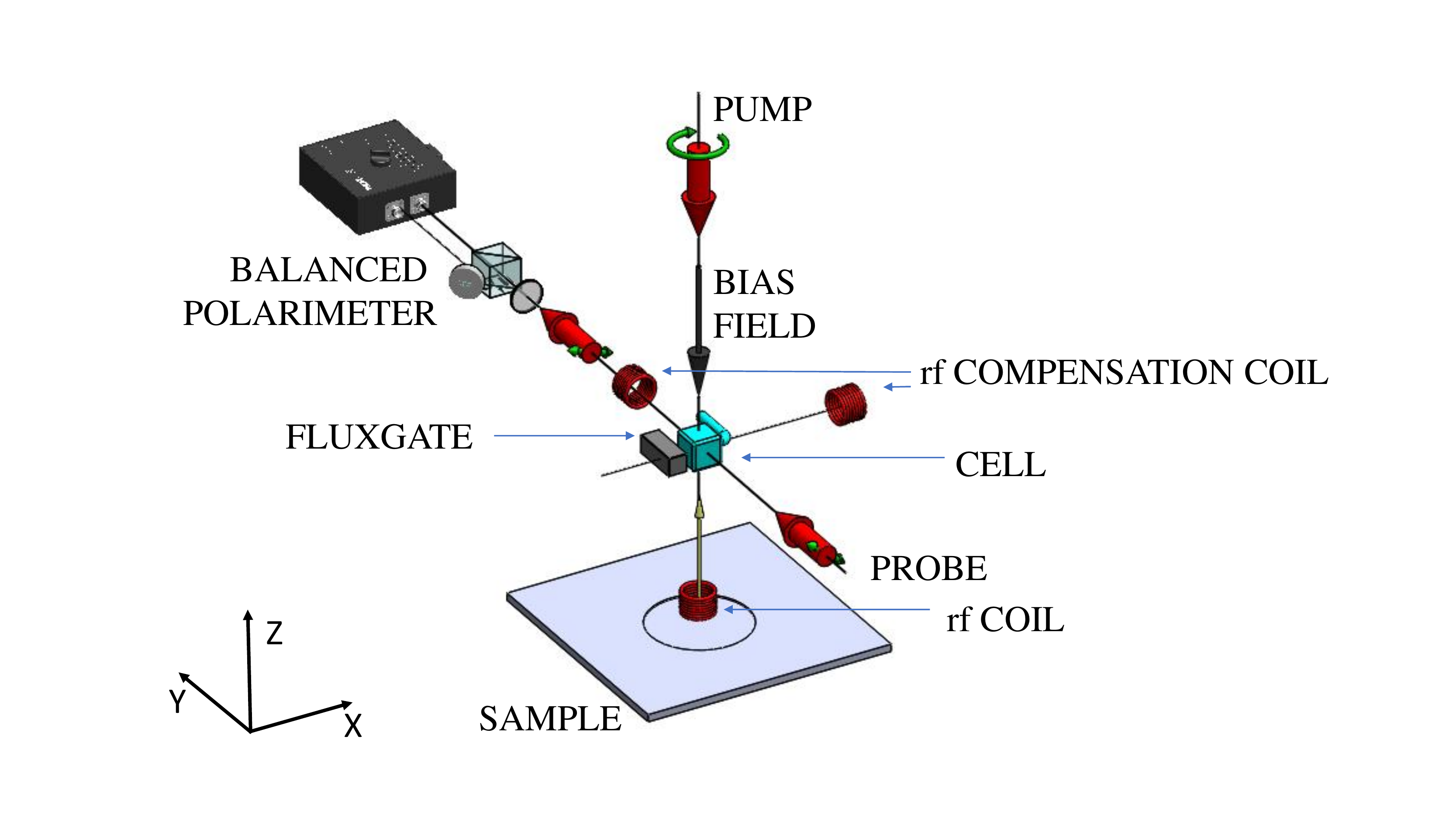}
\caption{Modified experimental setup with $\vec{B}_{bias}$ along $\hat{z}$. The pump laser beam orients the atomic vapour along the direction of the bias field. The horizontal components of the primary field are compensated in the vapour cell by a set of two rf coils oriented along the $\hat{x}$ and $\hat{y}$ directions.}\label{fig:Setup_vertical}
\end{figure} 

%\textbf{Novel geometry} 
The atomic magnetometer evaluates the oscillating magnetic field strength through the measurement of the atomic Zeeman coherence amplitude produced by this field in the atomic vapour polarized along the direction of the static magnetic field, $\vec{B}_\mathrm{bias}$ \cite{Chalupczak2018}. Since only magnetic fields oscillating orthogonally to the $\vec{B}_\mathrm{bias}$ direction can generate atomic coherences, the magnetometer is insensitive to rf fields along $\vec{B}_\mathrm{bias}$. For $\vec{B}_\mathrm{bias}$ aligned along $\hat{z}$, this property of the atomic magnetometer is equivalent to the first part of the compensation condition, $B'_z + b_\mathrm{z, max} = 0$, i.e. the absence of the z-component of the resultant rf field in the magnetometer signal. Figure \ref{fig:Setup_vertical} shows the modified measurement setup used in the demonstration of this type of compensation. The static bias magnetic field is directed along $\hat{z}$ and set to the same strength used in previous measurements (equivalent to Larmor frequency about $\SI{12.6}{\kilo\hertz}$). The pump laser beam is aligned along the bias magnetic field. Instrumentation includes a set of rf coils compensating the horizontal components of the primary field. The position of these compensating coils is varied such that the rf spectroscopy signal is minimised without the sample, in other words, to compensate the horizontal components of the primary field ($B'_x=0$ and $B'_y=0$). In this modified configuration the amplitude (R=$\sqrt{(b_x)^2+( b_y)^2}$) and the phase ($\phi = \arctan\frac{ b_x}{ b_y}$)  of the rf spectroscopy signal reflects variations in the amplitude and phase of the horizontal components of the secondary field.

\begin{figure}[htbp]
\includegraphics[width=\columnwidth]{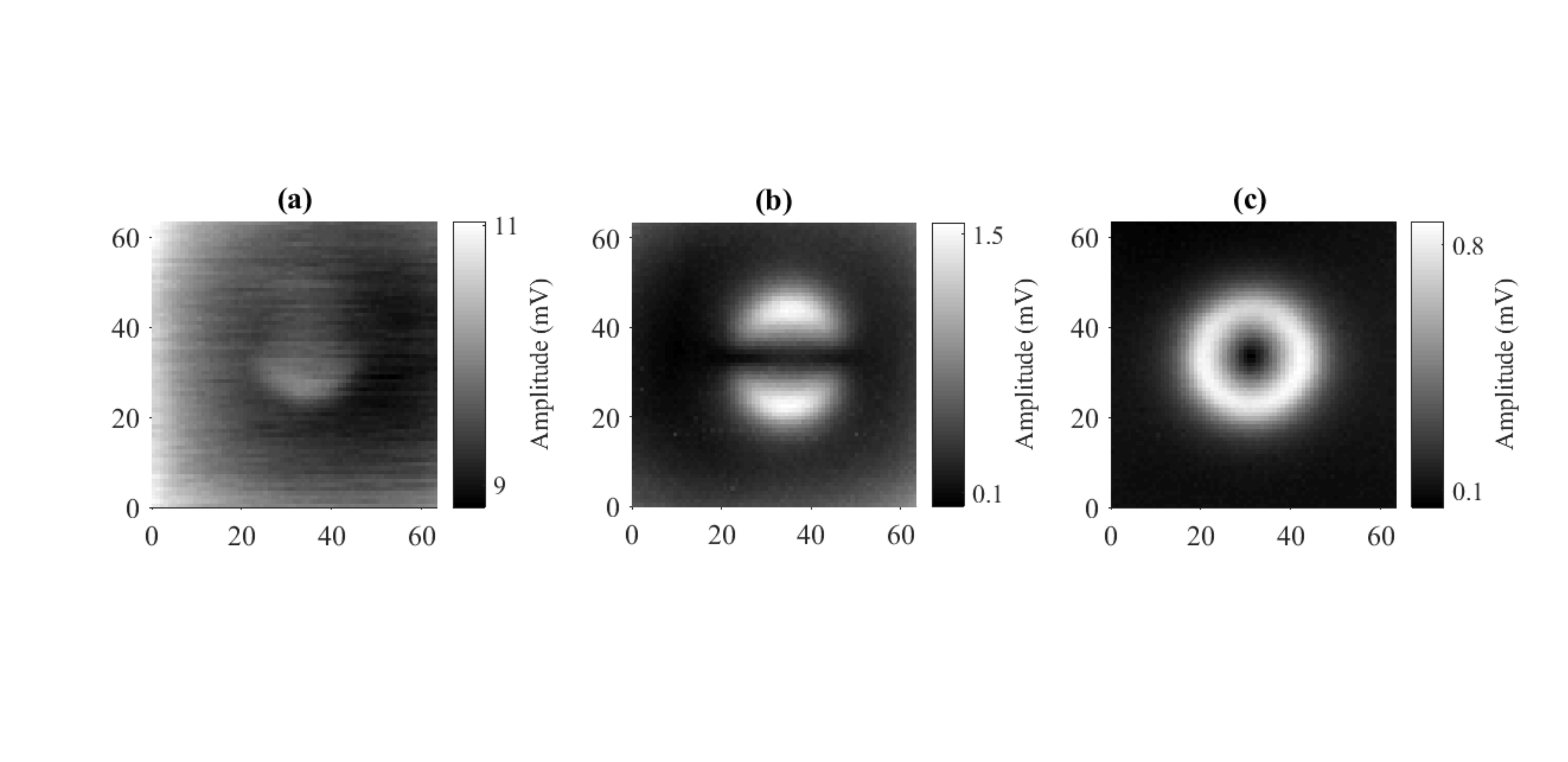}
\caption{The measured change of the amplitude of the rf spectroscopy signal over a $64\times\SI{64}{\milli\meter\squared}$ area of a \SI{6}{\milli\meter} thick aluminum plate, containing a \SI{24}{\milli\meter} diameter recess that is \SI{2.4}{\milli\meter} deep, recorded with three measurement configurations: (a) without rf field compensation, (b) with compensation performed with two rf coils  (Fig. \ref{fig:Setup}), (c) and with a rotated bias magnetic field and compensation coils (Fig. \ref{fig:Setup_vertical}).} \label{fig:Images}
\end{figure}

Figure \ref{fig:Images} illustrates the benefits of, and differences between, the two compensation schemes discussed. It shows the images of a $64\times\SI{64}{\milli\meter\squared}$ area of a \SI{6}{\milli\meter} thick aluminium plate containing a \SI{24}{\milli\meter} diameter recess that is \SI{2.4}{\milli\meter} deep recorded in three different configurations: (a) without compensation, (b) with compensation performed with two rf coils, and (c) with a rotated bias magnetic field and compensation coils. The images represent the change of the amplitude of the rf spectroscopy signal. As mentioned before, for the uncompensated case (a) the recorded profile shows a variation of the vertical component of the secondary field. In the compensated cases, the images show the $\hat{x}$ horizontal component (b) and both horizontal components (c) of the secondary field. The difference in the symmetry of the images results from the change of the direction of the bias field. In the case shown in Fig. \ref{fig:Images} (b) $\vec{B}_\mathrm{bias}$ is directed along the $\hat{x}$ axis and therefore only signatures produced by the recess edges parallel to that direction are present in the recorded profile (in other words, the edge parallel to $\vec{B}_\mathrm{bias}$ produces an oscillating secondary field perpendicular to $\vec{B}_\mathrm{bias}$ that can be seen by the magnetometer). In the case shown in Fig. \ref{fig:Images} (c), $\vec{B}_\mathrm{bias}$ is directed along the $\hat{z}$ axis and the recorded profile shows the whole contour of the recess. We have calculated the amplitude contrast, $C_R$, as defined previously for the three images. The numbers confirm [(a) $C_R=0.04$; (b) $C_R=0.77$; (c) $C_R=0.79$] that implementation of the compensation schemes allows easier identification of the structural defects in the amplitude images.
In the case of the modified geometry, the phase of the rf spectroscopy signal shows a vortex centered on the recess. 
In the ideal case of the rf coil perpendicular to the surface of the sample, optimum compensation (i.e. defined by the compensation condition) could be achieved only with the proper geometry of the atomic magnetometer (i.e. without compensation coils). In the case of the rf coil tilted with respect to the surface normal, compensation with only the atomic magnetometer aligned along the primary field direction would result in mixing of the secondary field components. This will be manifested by asymmetries in the Fig. \ref{fig:Images} (c) image. In this case, the optimum compensation could be achieved by aligning the atomic magnetometer's insensitive axis orthogonal to the sample surface and implementation of compensation coils in the horizontal plane.

%\textbf{Conclusions.}
We have demonstrated two simple methods for improvement of the contrast in the images representing the variations of the amplitude and phase of the rf spectroscopy signal  recorded by the atomic magnetometer in eddy currents NDT measurement. The methods are based on compensation of the components of the resultant rf magnetic field monitored by the atomic magnetometer by either a set of coils or the geometry of the measurement. Reduction of the amplitude of the rf signal monitored by the atomic magnetometer through the compensation process does not compromise the ability for defect detection. On the contrary, monitoring of the signal phase in compensated configuration provides the option of clear ($180^{\circ}$ phase change) signature of the inhomogeneity.

\begin{acknowledgements}
%This work was funded by the Innovate UK Energy Game Changer programme (IUK 132437). 
\end{acknowledgements}


\begin{thebibliography}{10}
\bibitem{Footnote2} Similar arguments, as given below, could be used for objects with a magnetic permeability.
\bibitem{Footnote3} Only the $\hat y$ direction matters, the other two axis are effectively insensitive.
\end{thebibliography}
\end{document}